\documentclass[twocolumn,prl,showpacs,preprintnumbers,amsmath,amssymb,a4paper,fleqn]{revtex4-1}

\usepackage{dcolumn}
\usepackage{epsfig}
\usepackage{amssymb,amsmath,bm,graphicx,multirow,mathtools}
\usepackage{color}

\def\be{\begin{equation}}
\def\ee{\end{equation}}
\def\bea{\begin{eqnarray}}
\def\eea{\end{eqnarray}}
\def\beeg{\begin{align}}
\def\eeg{\end{align}}

\def\m{\mu}
\def\n{\nu}

\def\p{\partial}
\def\a{\alpha}
\def\b{\beta}
\def\t{\theta}

\def\11{1\hspace{-0.6em}1}
\def\22{1\hspace{-0.48em}1}

\def\pl{\p\hspace{-0.5em}^{^{^\leftarrow}}}
\def\pr{\p\hspace{-0.5em}^{^{^\rightarrow}}}

\def\olp{\overleftarrow{p}}
\def\orp{\overrightarrow{p}}
\def\mhorp{\overrightarrow{\hat{\mathbf{p}}}}
\def\mholp{\overleftarrow{\hat{\mathbf{p}}}}
\def\olpp{\overleftarrow{\p}}
\def\orpp{\overrightarrow{\p}}

\def\wl{\overleftarrow{\hat{\mathbf{w}}}}
\def\wlr{\overrightarrow{\hat{\mathbf{w}}}}
\def\xx{\mathbf{x}}
\def\yy{\mathbf{y}}
\def\zz{\mathbf{z}}

\def\pph{\hat{\mathbf{p}}}
\def\xh{\hat{x}}
\def\yh{\hat{y}}
\def\zh{\hat{z}}
\def\xxh{\hat{\xx}}
\def\yyh{\hat{\yy}}
\def\zzh{\hat{\zz}}
\def\lth{\overleftarrow{\hat{\mathbf{T}}}}
\def\rth{\overrightarrow{\hat{\mathbf{T}}}}
\def\o{\Omega}

\def\i{\imath}

\def\mfl{\mathbf{\Lambda}}

\def\bb{\overleftrightarrow{\Delta}}

\begin{document}

\title{Generalizing Coordinate Non Commutativity}
\author{Abolfazl\ Jafari\footnote{jafari-ab@sci.sku.ac.ir}}
\affiliation{Department of Physics, Faculty of Science,
Shahrekord University, P. O. Box 115, Shahrekord, Iran}
\date{\today }

\begin{abstract}

\textbf{Abstract:} 
In this paper,  
we establish and employ a local framework 
to the first order of
Riemann's curvature tensor in order to develop the corresponding coordinate non commutativity 
into general manifolds.
We also exploit a new translation of function
at the level of quantum mechanics to show that the final correlation result of the generalized non commutativity is a mixture of the Canonical and Quadratic formalisms and does not consist only of the Lie algebraic formalism.
The basic premise of this article is that the geometry of a four-dimensional pseudo Riemann manifold representing space time,
is homeomorphic to Minkowski space time.
\end{abstract}


\maketitle


\section{Introduction}

Continuous space time makes available short distances.
Today, there are indications
that at very short distances we might have to go beyond differential manifolds.
Nowadays, we can formulate the fundamental laws of physics, 
consisting of field theory, gauge field
and the theory of gravity on differentiable manifolds.
This
is only one of the several issues that we confront in relation to changes in physics
for very short distances\cite{witten, connes, wess, nekrasov}.
Physics data has forced us to admit of change in the
conception of space time for very short distances and introduce non commutative coordinates \cite{wess, nekrasov, szabo1, szabo2, jab1,jab2}.
We define a non commutative space by
replacing the local coordinates $y^\m$ of $R^{D+1}$ with hermitian operators $\hat{y}^\m$ obeying the commutation
relations:
\begin{eqnarray} \label{canonical commutation relation}
&&[\hat{y}^\m,\hat{y}^\n]_\star=\omega^{\m\n}(\yyh,t),
\end{eqnarray}
in addition to the standard relations: $[\hat{y}^\m,\hat{p}^\n]=\i\delta^{\m\n}$ and $[\hat{p}^\m,\hat{p}^\n]=0.$
Where
$\hat{y}$ and $\hat{p}$ stand for the coordinates and momentum operators, respectively. 
$\omega(\yyh,t)$ is a real, antisymmetric four dimensional matrix.
Here, the Latin and Greek indices run from 1 to 3 and 0 to 3, respectively.
An important special case of the non commutativity (without time contribution) is as follows \cite{witten, nekrasov, szabo, bertolami, jaf4}:
\begin{align}
\label{1}
[\yh^{k},\yh^{l}]_\star=\i\t^{kl},
\end{align}
\noindent where $\t^{kl}=\left(
              \begin{array}{cccc}
                0 & 0 & 0 & 0 \\
                0 & 0 & \t^{12} & \t^{13} \\
                0 & \t^{21} & 0  & \t^{23} \\
                0 & \t^{31} & \t^{32} & 0
              \end{array}
            \right)$ 
is a constant and real tensor.
Of course, there are famous structures of non commutativity such as: the Lie algebraic $[\yh^{k},\yh^{l}]_\star=\i\t^{kl}_m\yh^m$ and Quadratic formalisms $[\yh^{k},\yh^{l}]_\star=\i\t^{kl}_{mn}\yh^m\yh^n$ where $\t^{kl}_m$ and $\t^{kl}_{mn}$ are constant structures. 
The formulation of physical theories on non commutative spaces is constructed by a very simple role, namely,
replacing the ordinary products between quantities with a new product, the so-called
$\star$ product named by Moyal-Weyl \cite{wess, nekrasov, szabo},
\begin{align}\label{mw}
f(\yyh) g(\yyh)=f(\yy)\star g(\yy)=f(\yy){\rm e}^{\frac{\i}{2}\pl_{\ \ \m}\t^{\m\n}\pr_{\ \ \n}}
g(\yy),
\end{align}
where $\p_\a=\frac{\p}{\p y^\a}$.
We emphasize that the set of coordinates without the symbol hat such as $"\yy"$ are commutable coordinates.

In this article, the canonical non commutativity can readily be reformulated to apply to general pseudo Riemann manifolds. 
In order for the above ideas to materialize, it is necessary to employ localized homomorphism between pseudo Riemannian manifolds. 
Locally, the accessibility of $\omega^{ij}\rightarrow \t^{ij}$ follows from the theory of general relativity.
Assuming that the equivalence principle holds, special relativity is available in the presence of a gravitational field.
In fact, one can always construct local inertial frames at a given event belonging to space time,
in which free particles would move along straight lines.
In such frames,
the components of the metric tensor can be expanded in terms of Riemann's tensors.
If $\eta_{\m\n}=(-1,1,1,1)$, then the components of the metric and the relevant affine connections (Christoffel multipliers) are given by,
\begin{eqnarray}\label{metric}
g_{\a\b}=\eta_{\a\b}+\frac{1}{3} R_{\a\m\b\n}y^\m y^\n,
\end{eqnarray}
and
\begin{eqnarray} \label{Affine}
\Gamma^{\m}_{\a\b}=\frac{1}{3}\eta^{\m\n}(R_{\b\epsilon\n\a}+R_{\n\b\a\epsilon})y^\epsilon,
\end{eqnarray}
where raising of the localized Lorentz indices are done with 
Eqs.(\ref{metric})\cite{parker1, parker2}.
The coordinate system of such a frame are called \textit{Riemann's normal coordinates} \cite{nestrov, misner, weber1, mare, inverno}.
It can be seen that up to the first order of $\t$, Eq.(\ref{metric}) and Eq.(\ref{Affine}) will remain unchanged for a non commutative framework.
A common tool for all approximated methods is to work with local coordinates.
That is, the validity of the above statements is limited to the non covariant observer.

\section{Presentation of the theory}
In this article, we suppose that all the employed functions are smooth and $C^\infty$
and the hat symbol is used for the non commutative coordinates and operators.
Here, we can represent an unusual interpretation of the basic vectors by
$$<y\mid{\rm e}^{-\i\ a\hat{p}_y}=:<\tilde{y}\mid$$
Therefore, at the quantum mechanical level, the basic vectors can have an unusual operational role.
The main purpose of our work is to investigate   
the validity of $n$-dimensional 
translation operators $\hat{\mathbf{T}}$.
The compact form of a well-defined set of the second class of translators will be (Ref.\cite{DeWitt}),  
\begin{align}\label{qd-taylor}
\hat{\mathbf{T}}(\mfl;\o)={\rm e}^{-\i\mfl_\o\cdot\pph}.
\end{align}
$\o$ is a somewhere where $\mfl$ as a function of position is small.
Particularly, if $\yy\in\p\o$, then $\Lambda(\yy)=0$. 
The evolution of the states occurs within the Heisenberg comprehensive formula;
\begin{align}\label{heis}
\hat{\mathbf{T}}^\dagger(\yy) \Psi(0)\hat{\mathbf{T}}(\yy)=\Psi(\yy),
\end{align}
in which $\hat{\mathbf{T}}(\yy)$ as a translation operator is very famous in quantum literature.
So, we need to introduce the Heisenberg picture which refers to the second class of translators (Eq.(\ref{qd-taylor})).
In the simple case, Eq.(\ref{qd-taylor}) has been assembled by:
\begin{align} \label{qtaylor}
\hat{\mathbf{T}}(\yy)={\rm e}^{-\i\yy\cdot\pph}.
\end{align} 

\section{Importation of star product}
In this section, we derive the $\star$-product for the non commutative framework in Minkowski space time.
In the special case of non commutativity, we only consider the case of $[\yh^1,\yh^2]=\i\t$.  
However, for the case of $\omega^{ij}=0$, we have $\hat{\mathbf{T}}_1(\yy) \hat{\mathbf{T}}_2(\yy)=\hat{\mathbf{T}}_2(\yy) \hat{\mathbf{T}}_1(\yy)$.
But, this is not always true and its validity is questioned by
Eq.(\ref{canonical commutation relation}).
So that, Eq.(\ref{1}) 
provides a set of arranged-ordered translators
Obviously, $\omega^{\m\n}$ is the cause of the arrange-ordered translators. 
Generally,
\begin{align}\label{cause}
\hat{\mathbf{T}}_1({\mathrm{\yyh}})\hat{\mathbf{T}}_2({\mathrm{\yyh}}) \neq \hat{\mathbf{T}}_2({\mathrm{\yyh}})\hat{\mathbf{T}}_1({\mathrm{\yyh}}).
\end{align}
Substituting, the unit operator in the non commutative framework, 
the key point of this paper appears which is to transfer the magnitude condition of $\Lambda(\yyh)$ to the limits of integration.
That is,
\begin{eqnarray}\label{arbitrary translation}
\hat{\mathbf{T}}(\mfl;\o)={\rm e}^{-\i\mfl_\o\cdot\pph}=
{\rm e}^{-\int_{\o}d\hat{v}\mid \yyh>\mfl(\yyh)\cdot\mathbf{\nabla}<\yyh\mid}.
\end{eqnarray}
It is clear that 
$\mfl(\yyh;\o)$ itself
consists of the necessary information of translation
and should still be very small, 
whereas on the right side 
of Eq.(\ref{arbitrary translation}), 
$\mfl(\yyh)$ does not consist of information
and the information of translation
has been delivered to the limits of integration.
In other words, the direct responsibility of $\mfl(\yyh;\o)$ can be reduced by integration.
Clearly, arrange-ordered translators satisfy the cumulative properties.
That is, Eq.(\ref{arbitrary translation}) is true.
We distinguish the momentum operators belonging to the Hilbert and Dual space, by introducing: $\olp=-\pph_{_{QM}}$ and $\orp=\pph_{_{QM}}$.
Now, In the neighborhood of $\o$, we can introduce:
\bea\label{right-left} && 
\wlr(\mfl;\o)=\mfl_\o\cdot\mhorp,
\nonumber\\&\mathrm{and}&\nonumber\\&&
\wl(\mfl;\o)=\mholp\cdot\mfl_\o,
\eea
Substituting matrix elements of $<\xxh\mid \orp_i\mid \yyh>=-\i\delta_{,\xh^i}{(\xxh-\yyh)}$ and $<\xxh\mid \olp_i\mid \yyh>=\i\delta_{,\yh^i}{(\xxh-\yyh)}$ into Eq.(\ref{right-left}), (\ref{right-left}) becomes
\bea \label{2} &&
<\xxh\mid\wlr(\mfl;\o)\mid \yyh>=-\i\Lambda^i(\xxh;\o)\delta_{,\xh^i}{(\xxh-\yyh)},\nonumber\\&&
<\xxh\mid\wl(\mfl;\o)\mid \yyh>=\i\delta_{,\yh^i}{(\xxh-\yyh)}\Lambda^i(\yyh;\o).
\eea
In this way, we show that the second class of the translators can be enabled to make Eq.(\ref{mw}).

\textbf{Derivation of the $\star$-product:}
In order to derive explicit expressions for the $\star$-product, we can refer to the Heisenberg picture. 
If, $\mathbf{\hat{T}}(\mfl;\o)={\rm e}^{-\i\wlr(\mfl;\o)}$, then
Eq.(\ref{heis}) and the translation operator provided that,
\begin{align}
\mathbf{\hat{T}}^\dagger(\yyh;\o) h(\yyh_0)\mathbf{\hat{T}}(\yyh;\o)\mathbf{\hat{T}}^\dagger(\yyh;\o) f(\yyh_0)\mathbf{\hat{T}}(\yyh;\o)
\cr
=
h(\yyh_0,\yyh)f(\yyh_0,\yyh)\mid_\o,
\end{align}
with assumption $\mfl=\yyh$ and the validity of the above statement is limited to $\o$.
Thus, every displacement
on the non commutative pseudo Riemann manifolds is strongly dependent on the commutation
relations given by Eq.(\ref{canonical commutation relation}).
Therefore, the advent of a non trivial operator is inevitable.
The central operator can be represented as,
\begin{align}
\diamond_{\mathrm{\o}}=\lth(\mfl;\o)\rth(\mfl;\o).
\end{align}
Specifically, we will follow "$\diamond_\o$" in the case of non commutativity which lies on the $xoy-$plane. 

\textbf{Lemma-} The $\diamond_\o$ enables us to create Eq.(\ref{mw}).

\textbf{Proof-}
We note that the second class of the translators up to the first order of $\t$, obey the equation: $({\rm e}^{-\i\wlr(\mfl;\o)})^\dagger={\rm e}^{\i\wl(\mfl;\o)}$
therefore, we can state the following:
\begin{align} \label{diamond}
\diamond_\o= {\rm e}^{\i\wl(\mfl;\o)} {\rm e}^{-\i\wlr(\mfl;\o)}.
\end{align}
By dropping the redundant indices, the above equation consists of: $[\wl,\wlr]=\wl\wlr-\wlr\wl\ ,$ in which 
the second term should to be clarified. 
In order to explain the second term, 
we can show,
\begin{align}
[\wl,\wlr]
=\olp_j[\Lambda^j,\Lambda^i]_\star\orp_i+R.S,
\end{align}
in which $\mathrm{R.\ S}$ is the sentences which do not contain $\t$.
Thus, Eq.(\ref{diamond}) becomes
\begin{align}
\label{msp2}
\diamond_\o&={\rm e}^{\i\wl(\mfl;\o)-\i\wlr(\mfl;\o)
+\frac{1}{2}[\wl(\mfl;\o),\wlr(\mfl;\o)]}\nonumber\\
&={\rm e}^{\Lambda^i_{\o,i}
+\frac{1}{2}\overleftarrow{\hat{p}}_j[\Lambda^j_\o,\Lambda^i_\o]_\star\overrightarrow{\hat{p}}_i+R.S}.
\end{align}
Subsequently, to obtain the $\star$-product, we can take $\Lambda^i_\o=\hat{y}^i;\ i=1,2$ and $\o$ as a neighborhood of the origin.
In this way, Eq.(\ref{msp2}) gives:
\begin{eqnarray}\label{msp}
\diamond_\o=
\exp\{\delta^i_i+\frac{1}{2}\int_\o\int_\o d^n\xh d^n\yh d^n\zh  \mid\xxh>
\cr\times
\delta_{,\xh^i}{(\xxh-\yyh)}
[\yh^i,\yh^j]\delta_{,\zh^j}{(\yyh-\zzh)}) <\zzh\mid+R.S\}.
\end{eqnarray}
But, for a small $\o$, $\o$ is locally homomorphic to the pseudo Riemann manifolds. 
This allows us to employ Eq.(\ref{1}).
Therefore, assuming $$\lim_{\o\rightarrow0}[\yh^i,\yh^j]_\star\sim\t^{ij}$$ 
and up to the first order of the small parameters, $\diamond_\o$
reduces to the $\star$-product and Eq.(\ref{msp}) will be an analog of the Moyal-Weyl mapping.
However, obtained inevitable operator contains of two different parts which one could contribute to star product.
Thus, up to the first order of $\t$, 
the initial star product; $\star_{\mathrm{initial}}$ (referred to $\diamond_\o(\t)$) is as follows:
\begin{align}\label{31}
\star_{\mathrm{initial}}=
{\rm e}^{\frac{\i}{2}\t^{ij}\int_\o\int_\o d^nx d^ny\
\mid\xx>\frac{\olpp}{\p x^i}\ \delta{(\xx-\yy)}\ \frac{\orpp}{\p y^j}\ <\yy\mid}.
\end{align}
Integration can be removed by refunding information to the transition function.
In this case and for a neighborhood of the origin, Eq.(\ref{31}) can be written in the momentum representation as the operator: 
\begin{align}
\star_{\mathrm{initial}}={\rm e}^{\frac{\i}{2}\olp_i\t^{ij}\orp_j}.
\end{align}
So,
our conclusions leads us to the canonical non commutativity,
\begin{align}
[\hat{y}^i,\hat{y}^j]_\star=\i\t^{ij}\hat{\11}.
\end{align}

\subsection{Another Minkowski space time}
We now generalize the previous process for a space time other than the Minkowski space time.
Since, in quantum mechanics, the time coordinate does not manifest in the role of operator,
"time" does not contribute to coordinate non commutativity.
However, from the Heisenberg picture,
we can introduce time dependent vectors: $\mid \yy;t>$.
We also assume that all the operators and vectors will be defined at time $"t"$.
This means that, with good approximation, we can suppose that the framework
is falling along a geodesic $\mathcal{G}$ of space time.
In Riemann's normal coordinates, each space-like hypersurface of constant $"t"$ is normal to this geodesic and contains
the set of space like geodesics normal to $\mathcal{G}$. 
The time $"t"$ of an event in a hypersurface is the proper time along $\mathcal{G}$ at which the every point intersects
the hypersurface \cite{parker1, parker2}.
Consequently, 
the unit operator is made in compliance with the principle of symmetrically:
\begin{align}
\hat{\11}(t)=\int d^{n-1} \yy \mid \yy;t> \sqrt{-g(\yy;t)} <\yy;t\mid.
\end{align}
Without loss of generality, we can replace $"\mid \yy,t>"$ with $"\mid y>"$ which is a symbolic notation.
In addition, by introducing $d^n y\ =d^{n-1}\yy\ dt\ \delta{(t-\acute{t})}$, the unit operator becomes:
\begin{align}
\hat{\11}=\int d^n y\ \mid y>\ \sqrt{-g(y)}\ <y\mid.
\end{align}
According to Ref.\cite{DeWitt} and in commutative algebra, the generalized momentum operator, $\hat{\Pi}$ is given by:
\begin{align}
<x\mid\hat{\Pi}_k\mid y>=
\cr
-\i\frac{\p}{\p x^{k}}\delta{(x,y)}-\frac{\i}{4}\Big{(}\frac{\p}{\p x^{k}}\ln{(-g(x))}\Big{)}
\delta{(x,y)},
\end{align}
with, 
\begin{align}\label{dell}
<x\mid y>=\delta{(x,y)}=\frac{\delta{(x-y)}}{\sqrt{-g(x)}}.
\end{align}
Eq.(\ref{dell}) is not always true and its validity is questioned by
Eq.(\ref{canonical commutation relation}).
For non commutative coordinates,
we can generalize the \text{rotation rule} as:
\begin{align}\label{deltafunction}
\sqrt{-g(\xh)}\ \delta{(\xh,\yh)}=\delta{(\xh-\yh)}=\delta{(\xh,\yh)}\ \sqrt{-g(\yh)}.
\end{align}
Also,
in commutative algebra, $\p_\m(-g(y))^{\frac{1}{2}}=(-g(y))^{\frac{1}{2}}\Gamma^\a_{\a \m}(y)$  
and the metric functions obey: $g(y)g^{-1}(y)=1$ \cite{kleinert}. 
One can see that the corresponding non commutative version of the metric functions obeys the same equation: $g(\yh)g^{-1}(\yh)=1$,
because we can ignore $(\Gamma^\a_{\a k})^2$ in $$g(\yh) g^{-1}(\yh)=g(y)g^{-1}(y)+\frac{\i}{2}\t^{ij}\Gamma^\a_{i\a}\Gamma^\b_{j\b}+...,$$
therefore, the metric functions will be commutable. 
Also, due to Eq.(\ref{metric}) and Eq.(\ref{Affine}), we can write
\begin{align}
(-g(\yh))^{\frac{1}{2}}_{,\m}=
-\frac{1}{6}(R_{\m k}-2R_{0\m0k})\yh^k:=\Gamma_\m(\yh),
\end{align}
Considering the commutators of the metric functions and Eq.(\ref{deltafunction}), we have
\bea
\delta_{,\yh^k}{(\xh,\yh)}=-\delta_{,\xh^k}{(\xh,\yh)}.
\eea
Our calculations conclude that: 
\bea
(-g(y))^{\frac{1}{4}}<y\mid\Pi_k=-\i\frac{\p}{\p y}<\bar{y}\mid,
\eea
where $<\bar{y}\mid=(-g(y))^{\frac{1}{4}}<y\mid$ and 
as well as form the left.
Now, we introduce the following non commutative version of the generalized momentum operator:
\bea\label{gm0}
&&\overrightarrow{\hat{\mathfrak{D}}}_k=\orp_k+\Delta_k\circ,
\nonumber\\&&
\overleftarrow{\hat{\mathfrak{D}}}_k=\olp_k+\circ \Delta_k,
\eea
which operate from the right and left.
By the matrix elements:
\bea \label{gm1}&&
<\bar{\xh}\mid\overrightarrow{\hat{\mathfrak{D}}}_k\mid \bar{\yh}>=-\i\delta_{,\xh^k}{(\xh,\yh)}+\Delta_k\circ,
\nonumber\\&&
<\bar{\xh}\mid\overleftarrow{\hat{\mathfrak{D}}}_k\mid\bar{\yh}>=\i\delta_{,\yh^k}{(\xh,\yh)}+\circ \Delta_k.
\eea
Such that, the effect of which on a typical function $\mathfrak{R}$ 
which can be chosen as a function of the spinorial field $\psi$, vector components $v_j$ or scalar function given by:
\begin{align} \label{gm2}
<\bar{\xh}\mid\overrightarrow{\hat{\mathfrak{D}}}_k\mid\mathfrak{R}>=-\i\mathfrak{R}_{,k}(\xh)+(\Delta_k,\mathfrak{R})(\xh),
\cr
<\mathfrak{R}\mid\overleftarrow{\hat{\mathfrak{D}}}_k\mid\bar{\yh}>=\i\bar{\mathfrak{R}}_{,k}(\yh)+(\bar{\mathfrak{R}},\Delta_k)(\yh).
\end{align}
Where $(\Delta_k,v_j)(\xh)=\i\Gamma^i_{kj}(\xh)v_i(\xh)$, $(\bar{v}_i,\Delta_k)(\yh)=-\i\bar{v}_i(\yh)\Gamma^i_{kj}(\yh)$, $(\Delta_k,\psi)(\xh)=\i\Gamma_k(\xh)\psi(\xh)$, $(\bar{\psi},\Delta_k)(\yh)=\bar{\psi}(\yh)\Gamma_k(\yh)$ and $<\psi\mid\yh>=\overline{\psi}(\yh)$.
Hence, it is possible to introduce additional structure: $(\Delta_k,\mathfrak{R})(\xh)=-eA_k(\xh)\mathfrak{R}(\xh)$ and $(\bar{\mathfrak{R}},\Delta_k)(\xh)=e\bar{\mathfrak{R}}(\xh)A_k(\xh)$ in which $\mathbf{A}$ is the vector potential.
The effect of $\ \Delta_k\circ\ $ or $\ \circ\Delta_k\ $ vanishes for any scalar function.
Also, it can be easily shown that:
\bea
\overrightarrow{\hat{\mathfrak{D}}}_k=\overleftarrow{\hat{\mathfrak{D}}}_k+\circ\bb_k\circ.
\eea
where $\circ\bb_k\circ=-\circ\Delta_k +\Delta_k\circ$.
Now, we can refer to the Heisenberg picture. 
Similar to the previous section, the advent of a non trivial operator is still inevitable, i.e.
\bea\label{msg22}
\diamond_\o&=&\lth(\mfl;\o)\rth(\mfl;\o)
\nonumber\\&=&
{\rm e}^{\i\wl(\mfl;\o)-\i\wlr(\mfl;\o)
+\frac{1}{2}[\wl(\mfl;\o),\wlr(\mfl;\o)]},
\eea
in which $\wlr(\mfl;\o)=\mfl_\o.\overrightarrow{\hat{\mathfrak{D}}}$ and $\wl(\mfl;\o)=\overleftarrow{\hat{\mathfrak{D}}}.\mfl_\o$.
In order to explain the latest term and for the sake of simplicity, we denote $\p_{\yh^k}$ by $\p_k$ (it is used exclusively for $\yh$ as a middle coordinate) and have
$\frac{\p}{\p_{\yh^\m}}\Lambda=\Lambda_{,\m}$.
So, we can use from: $[\Lambda,\overrightarrow{\hat{\mathfrak{D}}}_k]=
\i\Lambda_{,k}-(\Delta_k,\Lambda)$, $[\overleftarrow{\hat{\mathfrak{D}}}_k,\Lambda]=
-\i\Lambda_{,k}-(\Lambda,\Delta_k)$,
and 
\begin{align}
[\overleftarrow{\hat{\mathfrak{D}}}_i,\overrightarrow{\hat{\mathfrak{D}}}_j]=
\cr
[-p_i+\circ\Delta_i,p_j+\Delta_j\circ]=\i\Delta_{j,i}\circ+\i\circ\Delta_{i,j},
\end{align}
to calculate
\begin{align}\label{gcrr}
[\wl(\mfl;\o),\wlr(\mfl;\o)]=
\cr
\overleftarrow{\hat{\mathfrak{D}}}_{i}[\Lambda^i_\o,\Lambda^j_\o]_\star\overrightarrow{\hat{\mathfrak{D}}}_j+[\overleftarrow{\hat{\mathfrak{D}}}_i,\Lambda^j_\o]\Lambda^i_\o\overrightarrow{\hat{\mathfrak{D}}}_j+\Lambda^j_\o\overleftarrow{\hat{\mathfrak{D}}}_i[\Lambda^i_\o,\overrightarrow{\hat{\mathfrak{D}}}_j]
\cr
+\Lambda^j_\o[\overleftarrow{\hat{\mathfrak{D}}}_i,\overrightarrow{\hat{\mathfrak{D}}}_j]\Lambda^i_\o.
\end{align}
Thus, only the $\overleftarrow{\hat{\mathfrak{D}}}_{i}[\Lambda^i_\o,\Lambda^j_\o]_\star\overrightarrow{\hat{\mathfrak{D}}}_j$ term of Eq.(\ref{msg22}) can contribute to the non commutativity and we can easily verify that the other terms contain no the effect of non commutativity.
Eq.(\ref{msg22}) must be satisfied by Moyal-Weyl mapping, in order to achieve the non commutativity of Minkowski space time. 
We are now ready to develop the Moyal-Weyl mapping.
However, only one of the sentences that appear in Eq.(\ref{msg22}) plays a role in the star product.
Substituting the unit operator in Eq.(\ref{msg22}),
the extended product; $\star_{\mathrm{ext}}$ (referred to $\diamond_\o(\t)$) becomes: 
\begin{align}\label{star}
\star_{\mathrm{ext}}={\rm e}^{\lim_{{\o\rightarrow0}}\int_\o\int_\o d^n\xh\ d^n\zh\ \mid \bar{\xh}>\hat{Q}(\xh,\zh) <\bar{\zh}\mid},
\end{align}
up to terms of order $\t$.
Where,
\bea\label{gmw}
\hat{Q}(\xh,\zh)\cr=
\frac{1}{2}
\int_\o d^n\yh\ \ 
\overleftarrow{\hat{\mathfrak{D}}}_i(\xh,\yh)[\Lambda^i(\yh),\Lambda^j(\yh)]_\star
\overrightarrow{\hat{\mathfrak{D}}}_j(\yh,\zh),
\eea
where, $\overleftarrow{\hat{\mathfrak{D}}}_i(\xh,\yh)$ and $\overrightarrow{\hat{\mathfrak{D}}}_j(\yh,\zh)$ are given by Eqs.(\ref{gm0}) and (\ref{gm1}).
The integration of Eq.(\ref{star}) can be removed by restoring the transition's information and we must be careful to doing this.
So that Eq.(\ref{star}) becomes:
\begin{align}\label{gmw2}
\star_{\mathrm{ext}}=
{\rm e}^{\frac{1}{2}\overleftarrow{\hat{\mathfrak{D}}}_i[\Lambda^i,\Lambda^j]_\star\overrightarrow{\hat{\mathfrak{D}}}_j}.
\end{align}

\section{Locality}
Due to the principles of general relativity,
special relativity is possible on tangent space time.
So, up to the first order of the Riemann curvature tensors, one finds
\begin{align}
\omega^{ij}(\yh)\rightarrow\t^{ij}+\tilde{\t}^{ij}(\yh),
\end{align}
where, $\t$ is a constant part and $\tilde{\t}(\yh)$ is a coordinate dependent part of non commutativity.
The second term includes the first order of the Riemann curvature tensors\cite{misner, parker1, parker2}.
There is a homomorphism of local pseudo Riemann manifolds with Minkowski space time; so, we can employ Eq.(\ref{canonical commutation relation}).
Also, in the case study,
\begin{align}\label{plane}
[\yh^i,\yh^j]_\star=\i\t^{ij},
\end{align}
that is $\t^{0\m}=0$.
Therefore, Eq.(\ref{msg22}) becomes, 
$$\lth(\mfl;\o)\rth^\dagger(\mfl;\o)\mid_{[\yh^i,\yh^j]_\star=\i\t^{ij}}$$
Now, we can again set: $\Lambda^i=\yh^i$ and $\o$ as a neighbor of the origin is substituted in Eq.(\ref{gmw}), the Moyal-Weyl mapping becomes available
on other than Minkowski space time:
\begin{align}\label{Moyal-Weyl mapping}
A(\yh)B(\yh)=A(y)\ {\rm e}^{\hat{Q}}\ B(y),
\end{align}
$\hat{Q}$ is given by Eq.(\ref{gmw}) or Eq.(\ref{gmw2}). 
According to Eq.(\ref{plane}), 
the generalized Moyal-Weyl mapping will be:
\begin{align}
\star_{\mathrm{ext}}={\rm e}^{\frac{\i}{2}\overleftarrow{\hat{\mathfrak{D}}}_i \t^{ij}\overrightarrow{\hat{\mathfrak{D}}}_j}.
\end{align}
Hence, in the presence of the electromagnetic fields; $A_\m$ and in the Minkowski space time, we have 
\begin{align}\label{latest nonn}
[\hat{y}^i,\hat{y}^j]_\star=
+\i\t^{ij}
+\frac{e}{2}\t^{ik}A_{k}\hat{y}^j
+\frac{e}{2}\t^{kj}\hat{y}^i A_{k}.
\end{align}
This paper summarized the principle of the above coupling.
Now, our calculations give us: 
\begin{align}\label{latest non}
[\hat{y}^i,\hat{y}^j]_\star=
+\i\t^{ij}
-\i\t^{ik}\ \Gamma^j_{kl}\hat{y}^l
-\i\hat{y}^l \Gamma^i_{lk}\t^{kj}.
\end{align}
for the vectors and so for the spinorial fields:
\begin{align}\label{latest non1}
\bar{\phi}(\yh){\rm e}^{\frac{\i}{2}(\i\olpp_i-\i\Gamma_i)\t^{ij}(-\i\orpp_j+\i\Gamma_j)}\psi(\yh)=
\cr
\frac{\i}{2}\bar{\phi}_{,i}\t^{ij}\psi_{,j}-\frac{\i}{2}\bar{\phi}_{,i}\t^{ij}\Gamma_j\psi-\frac{\i}{2}\bar{\phi}\Gamma_i\t^{ij}\psi_{,j}.
\end{align}
This result gives new product for the spinorial functions:
\bea\label{latest non 2}
\star_{\mathrm{ext}}^{\mathrm{spinors}}={\rm e}^{\frac{\i}{2}\olpp_i\ \t^{ij}\ \orpp_j-\frac{\i}{2}\overleftarrow{\22}\Gamma^\m_{\m\ i}\t^{ij}\orpp_j-\frac{\i}{2}\olpp_i\t^{ij}\Gamma^\m_{\m j}\overrightarrow{\22}}.\nonumber\\&&
\eea
Interestingly, 
\bea
\star_{\mathrm{ext}}^{\mathrm{scalars}}=\star_{\mathrm{initial}}.
\eea
According to the condition: $(\mathbf{U}\mathbf{V})_{;k}=\mathbf{U}_{;k}\mathbf{V}+\mathbf{U}\mathbf{V}_{;k}$,
we mention that the extended product will be an associative product.

\section{Conclusion}
At the level of quantum mechanics, we have found a way to generalize coordinate non commutativity to a general space time other than the Minkowski space time.
For the case of Eq.(\ref{plane}), 
we made the local operators which were arrange-ordered and satisfied the commutation relation of Eq.(\ref{cause}).
The translation information was entrusted into the range of integration, by substituting the unit operator in the second class of translators. 
Also, we showed that the new operator (named $"\diamond"$) can be made by employing the evolution operator (in the Heisenberg picture), homomorphism of local pseudo Riemann manifolds with Minkowski space time and the assumption of Eq.(\ref{plane}).
$"\diamond"$ enabled us to generalize coordinate non commutativity for a more general space time.
It can be seen that 
obtained coordinates non commutativity: $[\hat{y}^i,\hat{y}^j]_{\star_{\mathrm{ext}}}=+\i\t^{ij}-\i\t^{ik}\ \Gamma^j_{kl}\ \hat{y}^l-\i\yh^l\Gamma^i_{lk}\t^{kj}$ will be a mixture of the canonical and Quadratic formalisms. 
Further, we have generalized the Moyal-Weyl mapping to Eq.(\ref{Moyal-Weyl mapping})
which consists of the manifold structures.
We also showed that the obtained new product is an associative product.

\section{Acknowledgments}
The author thanks Shahrekord University for supporting this work with a research grant.
\newline

\end{document}